\documentclass[a4paper,11pt]{article}
\usepackage{pos}

\title{Towards a SM prediction for CP violation in charm}

\author[a]{Alexander Lenz}
\author*[a]{Maria Laura Piscopo}
\author[a]{Aleksey V.\ Rusov}

\affiliation[a]{Theoretische Teilchenphysik, Center for Particle Physics Siegen, Physik Department, Universit\"at Siegen, Walter-Flex-Str. 3, 57068 Siegen, Germany}

\emailAdd{alexander.lenz@uni-siegen.de}
\emailAdd{maria.piscopo@uni-siegen.de}
\emailAdd{rusov@physik.uni-siegen.de}

\abstract{We provide an overview of the current experimental and theoretical status of charm CP violation and discuss recent progress in obtaining a Standard Model prediction for $\Delta a_{\rm CP}^{\rm dir}$ using the framework of light-cone sum rules. Furthermore, we present new results for the ratios of the direct CP asymmetries and of the branching fractions for the modes $D^0 \to \pi^+ \pi^-$ and $D^0 \to K^+ K^-$.}

\FullConference{20th International Conference on B-Physics at Frontier Machines (Beauty2023)\\
 3-7 July, 2023\\
Clermont-Ferrand, France\\}

\begin{document}
\maketitle

\section{Introduction}
The charm sector offers a unique system for testing 
the Standard Model of particle physics~(SM), see Ref.~\cite{Lenz:2020awd} for a recent review. The peculiarities of charm are twofold. 
On the one hand, achieving precise theoretical predictions for charm observables is currently very challenging. This follows from the value of the charm quark mass which 
lies at the boundary between the heavy and the light quark regimes, such that the typical theoretical methods employed for the study of heavy hadrons might be less suitable or even inapplicable for the description of charmed systems. The behaviour of both the perturbative and power expansions becomes, in fact, a priori questionable as
\begin{equation}
\alpha_s(m_c) \sim 0.35\,, \qquad \frac{\Lambda_{\rm QCD}}{m_c} \sim  0.30\,.
\end{equation}
On the other hand, charmed hadrons provide essential complementary information with respect to
kaon- and $b$-physics, constituting, for instance, the only system to study meson-mixing in the up-quark sector.
Additionally, 
the sensitivity to potential new physics (NP) contributions is  high for charm observables, as pronounced cancellations often affect their SM predictions. The latter follow from the Glashow-Iliopoulos-Maiani~(GIM) mechanism due to $m_b, m_s, m_d \ll m_W$,
as well as from the size of the relevant elements of the Cabibbo-Kobayashi-Maskawa (CKM) matrix
i.e.\ $\lambda_q \equiv V_{cq}^* V_{uq}$, namely
\begin{equation}
\lambda_d = - 0.21874 + 2.51 \times 10^{-5} i\,,  \qquad \lambda_s = 0.21890+ 0.13 \times 10^{-5}i\,,
\end{equation}
\begin{equation}
\lambda_b = 6.3 \times 10^{-5} - 1.4 \times 10^{-4} i \,.
\end{equation}
In particular, having $\lambda_b$ the biggest relative imaginary part but being much smaller in magnitude compared to $\lambda_{d,s}$, the amount of CP violation in the charm sector is expected to be small in the SM. Testing this result against the experimental data, although theoretically difficult, is clearly a task of primary importance in order to strengthen the current understanding of the SM and search for NP.   
\section{Experimental status of CP violation in charm}
The observation of CP violation in the charm sector was made in 2019 by the LHCb Collaboration~\cite{LHCb:2019hro}, by measuring the difference of the time-integrated CP asymmetries in the $D^0 \to K^+ K^-$ and $D^0 \to \pi^+ \pi^-$ modes, that is $\Delta A_{\rm CP} \equiv A_{\rm CP} (K^+ K^-) - A_{\rm CP}(\pi^+ \pi^-)$. The corresponding difference of the direct CP asymmetries in the above channels turned out to be 
\begin{equation}
\Delta a_{\rm CP}^{\rm dir}\big|_{\rm exp} = (-15.7 \pm 2.9) \times 10^{-4}\,.
\label{eq:Delta_acp}
\end{equation}
Recently, also a measurement of the CP asymmetry in $D^0 \to K^+ K^-$ was published by the LHCb Collaboration~\cite{LHCb:2022lry}, which yields, when combined with the result in Eq.~\eqref{eq:Delta_acp}, the following values for the direct CP asymmetries in the two individual modes, namely
\begin{equation}
a_{\rm CP}^{\rm dir}(K^+ K^-)\big|_{\rm exp} = (7.7\pm 5.7)\times 10^{-4}\,, \qquad 
a_{\rm CP}^{\rm dir}(\pi^+ \pi^-)\big|_{\rm exp} =  (23.2\pm 6.1)\times 10^{-4}\,.
\label{eq:akk-apipi}
\end{equation}
While the result for $a_{\rm CP}^{\rm dir}(\pi^+\pi^-)$ provides the first evidence for CP violation in a specific $D$-meson decay, a clear theoretical interpretation of the measurements in Eqs.~\eqref{eq:Delta_acp}, \eqref{eq:akk-apipi} is currently still missing, particularly as the values of the individual CP asymmetries in Eq.~\eqref{eq:akk-apipi} would imply a surprisingly large breaking of the U-spin symmetry~\cite{Schacht:2022kuj}.
\section{Theoretical status of CP violation in charm}
Exclusive hadronic decays of charmed hadrons pose significant challenges for robust theoretical predictions and although several studies have been carried out in the literature, no unanimous conclusion has yet been reached on the origin of the experimental value of $\Delta a_{\rm CP}^{\rm dir}$.\\
Naive estimates, see e.g.~Ref.~\cite{Grossman:2006jg}, point towards a value of $\Delta a_{\rm CP}^{\rm dir}$ about an order of magnitude smaller than the one in Eq.~\eqref{eq:Delta_acp}. 
This result was confirmed in Refs.~\cite{Khodjamirian:2017zdu, Lenz:2023rlq} using the framework of light-cone sum rule~(LCSR) \cite{Balitsky:1989ry}, 
and analogous conclusions were also obtained in a recent study of final state interactions~\cite{Pich:2023kim}. Consequently, following these findings, several investigations of possible NP scenarios have been triggered in the effort to accommodate the experimental data~\cite{Chala:2019fdb}.\\
Furthermore, also SM interpretations of the experimental value of $\Delta A_{\rm CP}$ have been advanced. 
These include analyses based on U-spin relations, see e.g.\ Ref.~\cite{Grossman:2019xcj}, as well as studies of rescattering contributions~\cite{Schacht:2021jaz} and of final state interactions~\cite{Bediaga:2022sxw}. In particular, in Ref.~\cite{Schacht:2021jaz}, the possibility 
that nearby resonances, like the $ f_0(1710)$ or $f_0(1790)$, could lead to a large enhancement of the SM value of $\Delta A_{\rm CP}$, was pointed out. No sign of this effect, however, has been observed in the analysis of Ref.~\cite{Pich:2023kim}, and the latter work also indicated some inconsistencies in Ref.~\cite{Bediaga:2022sxw}. Finally, approaches based on topological diagram analyses have also been employed~\cite{Li:2019hho}, although these often rely on qualitative studies and do not provide a first principle determination.
\section{\boldmath Theory of the decays $D^0 \to \pi^+ \pi^-$ and $D^0\to K^+ K^-$ }
\begin{figure}[t]
    \begin{minipage}{0.48\textwidth}
    \centering
    \includegraphics[scale=0.43]{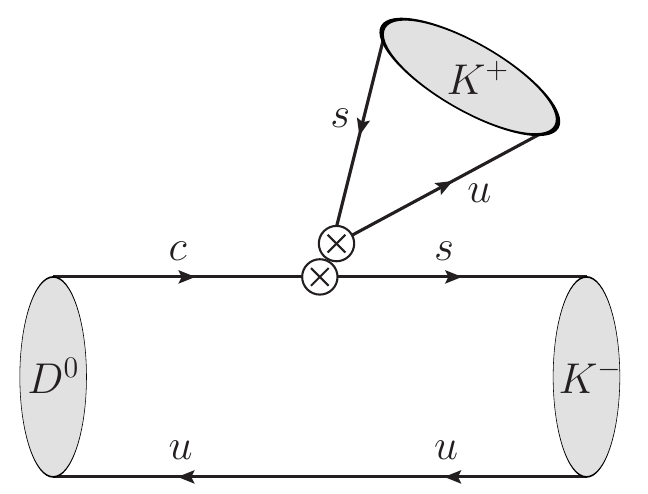}\\
    (a)
    \end{minipage}
    \begin{minipage}{0.48\textwidth}
    \centering
   \hspace*{-1cm} \includegraphics[scale=0.43]{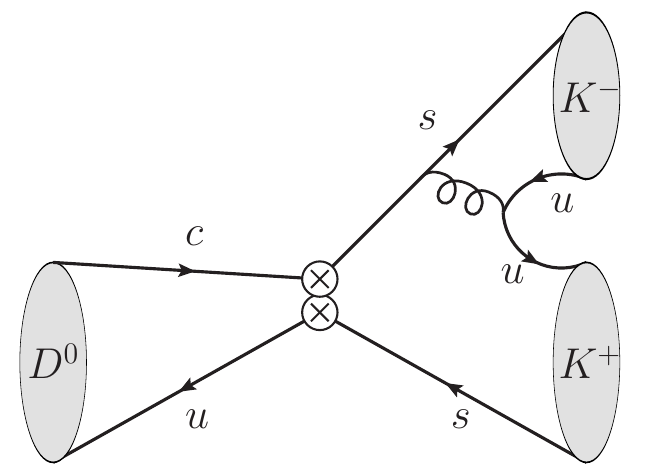}\\
    (b)
    \end{minipage}
   \begin{minipage}{0.48\textwidth}
    \centering
    \includegraphics[scale=0.43]{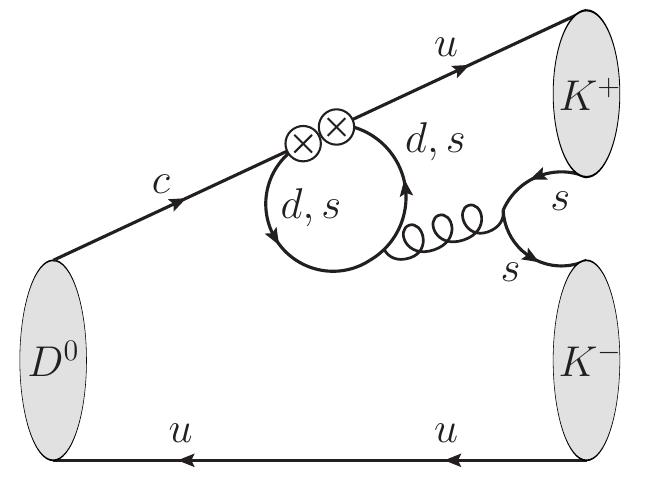}\\
    (c)
    \end{minipage}
    \begin{minipage}{0.48\textwidth}
    \centering
    \includegraphics[scale=0.43]{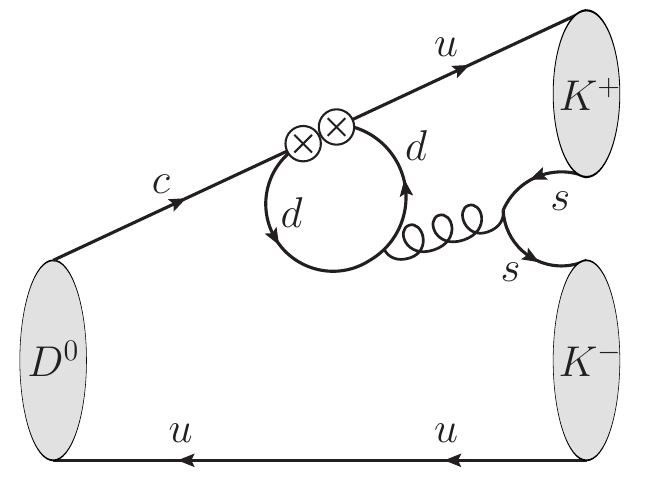}\\
    (d)
    \end{minipage}
    \caption{Examples of tree-level (a), exchange (b) and penguin (c) topologies contributing to ${\cal A}_{KK}$. Example of penguin topology contributing to ${\cal P}_{KK}$ (d). The corresponding diagrams for ${\cal A}_{\pi \pi}$ and ${\cal P}_{\pi \pi}$ can be obtained replacing $K \to \pi$, $s \leftrightarrow d$.}
    \label{fig:AKK-diagrams}
\end{figure}

Using the unitarity of the CKM matrix $\lambda_d + \lambda_s + \lambda_b = 0$, the amplitudes for the non-leptonic decays $D^0 \to \pi^+ \pi^-$ and $D^0 \to K^+ K^-$ can be recast in the form~\cite{Khodjamirian:2017zdu}
\begin{align}
    {\cal A}(D^0 \to \pi^+ \pi^-) & =  \lambda_d \, {\cal A}_{\pi \pi} \left[1 - \frac{\lambda_b}{\lambda_d} \frac{{\cal P}_{\pi\pi}}{{\cal A }_{\pi \pi}} \right] \,,
    \label{eq:A-Dpipi}
    \\[2mm]
   {\cal A}(D^0 \to K^+ K^-) &= \lambda_s \, {\cal A}_{K K} \left[1 - \frac{\lambda_b}{\lambda_s} \frac{{\cal P}_{KK}}{{\cal A }_{KK}} \right]\,,
    \label{eq:A-DKK}
\end{align}
by singling out, respectively, the contribution due to the CKM dominant combination $\lambda_{d,s} $ from that of the strongly suppressed factor $\lambda_b$, with the definitions 
\begin{align}
{\cal A}_{\pi\pi} & = \langle \pi^+ \pi^-| {\cal O}^d| D^0 \rangle - \langle \pi^+ \pi^-| {\cal O}^s| D^0 \rangle\,,
 \label{eq:Apipi}
\\[2mm]
{\cal A}_{KK} & = \langle K^+ K^-| {\cal O}^s| D^0 \rangle - \langle K^+ K^-| {\cal O}^d| D^0 \rangle\,,
 \label{eq:Akk}
\end{align}
and 
\begin{equation}
    {\cal P}_{\pi\pi} = \langle \pi^+ \pi^-| {\cal O}^s| D^0 \rangle\,, \qquad   {\cal P}_{KK} = \langle K^+ K^-| {\cal O}^d| D^0 \rangle\,.
    \label{eq:P}
\end{equation}
In Eqs.~\eqref{eq:Apipi} - \eqref{eq:P}, the notation ${\cal O}^q \equiv - (G_F/\sqrt 2)\sum_{i=1,2}C_i O_i^{q}$ is used, where $O^q_{1} = (\bar q^i \Gamma_\mu c^i) (\bar u^j \Gamma^\mu q^j)$ and $O^q_2 = (\bar q^i \Gamma_\mu c^j) (\bar u^j \Gamma^\mu q^i)$, with $q = d, s$, denote the current-current operators in the weak effective Hamiltonian describing the charm-quark transitions $c \to q \bar q u$~\cite{Buchalla:1995vs}, and $C_{1,2}$ are the corresponding Wilson coefficients.
The leading CKM amplitudes ${\cal A}_{\pi\pi}$, ${\cal A}_{KK}$ in Eqs.~\eqref{eq:A-Dpipi}, \eqref{eq:A-DKK} receive contributions from color-allowed tree-level, exchange and penguin topologies, whereas only the penguin topology can contribute to ${\cal P}_{\pi\pi}$, ${\cal P}_{KK}$, cf.\  Fig.~\ref{fig:AKK-diagrams}.
\\
Considering for simplicity only the decay $D^0 \to K^+ K^-$, the corresponding branching fraction reads
\begin{equation}
   {\cal B}(D^0 \to K^+ K^-) \propto|\lambda_s|^2 |{\cal A}_{KK}|^2 \left|1 - \frac{\lambda_b}{\lambda_s} \frac{{\cal P}_{KK}}{{\cal A}_{KK}} \right|^2 \,,
   \label{eq:BR-KK}
 \end{equation}
up to phase-space and normalisation factors.
Similarly, the direct CP asymmetry, defined as
\begin{equation}
a_{\rm CP}^{\rm dir}(f) \equiv \frac{
\Gamma (D^0 \to f) - \Gamma (\overline{D}^0 \to \bar f)
}
{
\Gamma (D^0 \to f) + \Gamma (\overline {D}^0  \to \bar f)
}\,,
\end{equation}
becomes
\begin{equation}
    a_{\rm CP}^{\rm dir}(K^+ K^-) = - \frac{\displaystyle 2 \left|\frac{\lambda_b}{\lambda_s}  \right| \sin{\gamma} \left| \frac{{\cal P}_{KK}}{{\cal A}_{KK}} \right| \sin{\phi}_{KK} }{\displaystyle 1-2 \left|\frac{\lambda_b}{\lambda_s} \right| \cos{\gamma} \left| \frac{{\cal P}_{KK}}{{\cal A}_{KK}} \right| \cos{\phi}_{KK}  + \left|\frac{\lambda_b}{\lambda_s} \right|^2  \left| \frac{{\cal P}_{KK}}{{\cal A}_{KK}} \right|^2  }\,,
    \label{eq:acp_dir}
\end{equation}
where we have defined the strong phase difference $\phi_{KK} \equiv \arg \left({\cal P}_{KK}/{\cal A}_{KK} \right)$, and introduced the angle $\gamma \equiv -\arg(\lambda_b/\lambda_s)$. 
Analogous expressions can be straightforwardly obtained for the mode $D^0 \to \pi^+ \pi^-$ by replacing $K K \to \pi \pi$, $\lambda_s \to \lambda_d$ and $\sin \gamma \to - \sin \gamma$ in Eqs.~\eqref{eq:BR-KK}, \eqref{eq:acp_dir}.\\ 
Taking into account the large hierarchy $\lambda_b/\lambda_{d,s} \ll 1$, it follows that the amplitudes ${\cal A}_{\pi\pi}$, ${\cal A}_{KK}$ give the dominant contribution to the branching fractions, i.e.\ 
\begin{equation}
 {\cal B}(D^0 \to \pi^+ \pi^-) \sim|\lambda_d|^2 |{\cal A}_{\pi \pi}|^2 \,,
\qquad 
{\cal B}(D^0 \to K^+ K^-)\sim  |\lambda_s|^2 |{\cal A}_{KK}|^2\,,
\label{eq:BR-D0-to-K-K}    
\end{equation}
whereas the direct CP asymmetries are driven only by the ratio of the penguin over the CKM leading amplitudes, that is
\begin{equation}
 a_{\rm CP}^{\rm dir}(\pi^+ \pi^-) \simeq 2 \left|\frac{\lambda_b}{\lambda_d}  \right| \sin{\gamma} \left| \frac{{\cal P}_{\pi\pi}}{{\cal A}_{\pi\pi}} \right| \sin \phi_{\pi\pi}  \,, 
\qquad    
a_{\rm CP}^{\rm dir}(K^+ K^-) \simeq  - 2 \left|\frac{\lambda_b}{\lambda_s}  \right| \sin{\gamma} \left| \frac{{\cal P}_{KK}}{{\cal A}_{KK}} \right| \sin \phi_{KK}\,.
    \label{ew:a_CP_appr}
\end{equation}
Finally, the above results, together with $|\lambda_d| \simeq |\lambda_s|$, yield the following expression for the difference of direct CP asymmetries $\Delta a_{\rm CP}^{\rm dir}$, namely
\begin{equation}
\Delta a_{\rm CP}^{\rm dir} \simeq - 2 \left|\frac{\lambda_b}{\lambda_s} \right| \sin{\gamma} 
\left(\left| \frac{{\cal P}_{KK}}{{\cal A}_{KK}} \right| \sin \phi_{KK} 
+ \left| \frac{{\cal P}_{\pi\pi}}{{\cal A}_{\pi\pi}} \right| \sin \phi_{\pi\pi}\right) \,.
\label{eq:Delta-ACP}
\end{equation}  
\section{\boldmath Determination of $\Delta a_{\rm CP}^{\rm dir}$ within LCSR}
A first computation of the penguin amplitudes ${\cal P}_{\pi\pi}$, ${\cal P}_{KK}$ was performed in Ref.~\cite{Khodjamirian:2017zdu} using the framework of LCSR with, respectively, pion and kaon light-cone distribution amplitudes~(LCDAs), and following previous studies for the $B \to \pi \pi$ decay~\cite{Khodjamirian:2000mi}. 
The values of $|{\cal A}_{\pi \pi}|$ and $|{\cal A}_{KK}|$ needed to determine the direct CP asymmetries were instead extracted, taking into account the relations in Eq.~(\ref{eq:BR-D0-to-K-K}), from the precise experimental data on the branching ratios~\cite{ParticleDataGroup:2022pth}
\begin{align}
{\cal B} (D^0 \to \pi^+ \pi^-)\big|_{\rm exp}  & =  (1.454 \pm 0.024) 
\times 10^{-3} \, ,
\label{eq:Br_exp-pipi}\\[2mm]
{\cal B} (D^0 \to K^+ K^-)\big|_{\rm exp}    & =  (4.08 \pm 0.06) 
\times 10^{-3} \,.
\label{eq:Br_exp-KK}
\end{align}
The authors of Ref.~\cite{Khodjamirian:2017zdu} obtained the following SM bound for the difference of CP asymmetries
\begin{equation}
|\Delta a_{\rm CP}^{\rm dir}|_{\rm SM} \leq 2.3 \times 10^{-4}\,,
\label{eq:Delta_acp_KP}
\end{equation}
which is about a factor of 6 lower than the experimental value in Eq.~\eqref{eq:Delta_acp}.
Recently, a study of the leading decay amplitudes ${\cal A}_{\pi \pi}$, ${\cal A}_{KK}$ has been performed in Ref.~\cite{Lenz:2023rlq}, where the corresponding tree-level matrix elements have also been determined using LCSR with pion and kaon LCDAs, see also Ref.~\cite{Piscopo:2023opf} for more details on the general framework. Specifically, from naive power counting, Eqs.~\eqref{eq:Apipi}, \eqref{eq:Akk} can be expressed as
\begin{align}
    {\cal A}_{\pi \pi} &=  \langle \pi^+ \pi^-| {\cal O}^d| D^0 \rangle \Bigl|_{\rm tree} + \, {\cal O}(\alpha_s) + {\cal O}(1/m_c)\,,
    \\[2mm]
    {\cal A}_{K K} &=  \langle K^+ K^-| {\cal O}^s| D^0 \rangle \Bigl|_{\rm tree} + \, {\cal O}(\alpha_s) + {\cal O}(1/m_c)\,,
\end{align}
retaining the dominant contribution due to the tree-level amplitude and neglecting sub-leading diagrams due to both hard and soft QCD corrections. In this approximation it follows that 
\begin{align}
{\cal A_{\pi \pi}} &\simeq -\frac{G_F}{\sqrt 2}  \left(C_1 + \frac{C_2}{3} \right)    
\langle \pi^+ \pi^- | O_1^{d} | D^0 \rangle \Bigl|_{\rm tree}\,,
\label{eq:Apipi-appr}
\\[2mm]
{\cal A}_{K K} &\simeq -\frac{G_F}{\sqrt 2}   \left(C_1 + \frac{C_2}{3} \right)    
\langle K^+ K^- | O_1^s | D^0 \rangle \Bigl|_{\rm tree}\,.
\label{eq:AKK-appr}
\end{align}
A first estimate of the matrix elements in Eqs.~\eqref{eq:Apipi-appr}, \eqref{eq:AKK-appr} can be derived using the naive QCD factorisation approximation, which, surprisingly, already yields values for the branching fractions in very good agreement with the experimental data~\cite{Lenz:2023rlq}. Furthermore, the computation of the tree-level matrix elements within the framework of LCSR gives~\cite{Lenz:2023rlq} 
\begin{align}
{\cal B} (D^0 \to \, \pi^+ \, \pi^-)\bigl|_{\rm LCSR}  & =  \left(1.40^{+1.53}_{-1.06}\right) \times 10^{-3}\,,
\label{eq:BR-D0-to-pi-pi-LCSR}\\[2mm]
{\cal B} (D^0 \to K^+ K^-)\bigl|_{\rm LCSR}  & =  \left(3.67^{+3.90}_{-2.69}\right) \times 10^{-3}\,,
\label{eq:BR-D0-to-K-K-LCSR}
\end{align}
where the central values again agree very well with the data, however, the uncertainties are large and mostly follow from a conservative treatment of missing contributions. Importantly, these results do not indicate any sign of potential large enhancement due to subleading topologies. On the other hand, in the ratio of branching fractions many theoretical uncertainties cancel, leading, after taking into account correlations due to common inputs, to the significantly more precise prediction
\begin{equation}
\frac{\displaystyle {\cal B} (D^0 \to \, K^+ \,K^-)}{\displaystyle {\cal B} (D^0 \to \pi^+ \pi^-)}\Bigg|_{\rm LCSR} = 2.63 \pm 0.86 \,,
\end{equation}
which perfectly reproduces the observed size of $SU(3)_F$ breaking in the two modes, namely
\begin{equation}
\frac{\displaystyle {\cal B} (D^0 \to \, K^+ \,K^-)}{\displaystyle {\cal B} (D^0 \to \pi^+ \pi^-)}\Bigg|_{\rm exp} = 2.81 \pm 0.06 \,.
\end{equation}
Combining the LCSR results for the penguin and tree-level amplitudes, as computed in Ref.~\cite{Khodjamirian:2017zdu} and Ref.~\cite{Lenz:2023rlq}, respectively, then yields the following value for the ratio of the two direct CP asymmetries defined in Eq.~\eqref{ew:a_CP_appr}, i.e.
\begin{equation}
 \frac{\displaystyle a_{\rm CP}^{\rm dir}(K^+ K^-)}{ \displaystyle a_{\rm CP}^{\rm dir}(\pi^+ \pi^-)} \Bigg|_{\rm LCSR}= - 0.65^{+ 0.09}_{- 0.08}\,,
 \label{eq:a_cp_ratio_LCSR}
\end{equation}
where we have used the estimates of the strong phases as obtained in Ref.~\cite{Khodjamirian:2017zdu}, since the sensitivity to potential missing contributions not yet accounted in the LCSR result is expected to be softened in the ratio $\sin \phi_{KK}/\sin \phi_{\pi \pi}$. The value in Eq.~\eqref{eq:a_cp_ratio_LCSR} is well consistent with -1, the result that would be obtained in the limit of exact U-spin symmetry, and must be compared with the corresponding experimental ratio
\begin{equation}
\frac{\displaystyle a_{\rm CP}^{\rm dir}(K^+ K^-)}{ \displaystyle a_{\rm CP}^{\rm dir}(\pi^+ \pi^-)} \Bigg|_{\rm exp} = 0.33^{+0.45}_{-0.26}  \,.
 \label{eq:a_cp_ratio_exp}
\end{equation}
Finally, allowing for arbitrary strong phase differences, that is varying both $\sin \phi_{\pi \pi}$ and $\sin \phi_{KK}$ from -1 to 1, the bound on $\Delta a_{\rm CP}^{\rm dir}$ obtained entirely using LCSR, reads~\cite{Lenz:2023rlq}
\begin{equation}
|\Delta a_{\rm CP}^{\rm dir}|_{\rm LCSR} \leq 2.4 \times 10^{-4} \,,
\label{eq:upperbound}
\end{equation}
in perfect agreement with the result in Eq.~\eqref{eq:Delta_acp_KP}, and again about a factor of 6 lower than the corresponding measurement.
\section{Conclusion}
We have briefly described the current experimental and theoretical status of charm CP violation and discussed recent progress obtained in the study of the hadronic decays $D^0\to K^+ K^-$ and $D^0 \to \pi^+ \pi^-$ using the framework of LCSR. In particular, we have shown that first steps towards a description of the corresponding branching fractions using this method yields very promising results~\footnote{See Section 6 of Ref.~\cite{Lenz:2023rlq} for a list of future improvements.}, and that LCSR leads to a bound on the value of $\Delta a_{\rm CP}^{\rm dir}$ in the SM which is about a factor of 6 lower than the experimental data. 
\section{Acknowledgements}
MLP is very grateful to the organisers of {\it Beauty2023} for the invitation and for creating a lively atmosphere rich of fruitful discussions. Moreover, AL and MLP would like to thank the participants of the recent LHCb school in Meinerzhagen for the interesting discussions and for proposing the computation of some of the observables presented in this manuscript.
The work of MLP is funded by the Deutsche Forschungsgemeinschaft (DFG, German Research Foundation) - project number 500314741. 

\end{document}